\documentclass{aanew}
\usepackage{psfig}

\def\bron{SAX~J1711.6-3808}
\def\1712{SAX~J1712.6-3739}
\def\ecs{erg~cm$^{-2}$s$^{-1}$}
\def\lum{erg~s$^{-1}$}

\begin{document}

\title{The nature of the X-ray transient \bron}
\titlerunning{\bron}
\authorrunning{J.J.M. in 't Zand, C.B. Markwardt et al.}

\author{J.J.M.~in~'t~Zand\inst{1,2}
 \and C.B. Markwardt\inst{3,4}
 \and A.~Bazzano\inst{5}
 \and M.~Cocchi\inst{5} 
 \and R.~Cornelisse\inst{1,2}
 \and J.~Heise\inst{2}
 \and E.~Kuulkers\inst{1,2}
 \and L.~Natalucci\inst{5}
 \and M.~Santos-Lleo\inst{6}
 \and J.~Swank\inst{4}
 \and P.~Ubertini\inst{5}
}

\offprints{J.J.M. in 't Zand, email {\tt jeanz@sron.nl}}

\institute{     Astronomical Institute, Utrecht University, P.O. Box 80000,
                NL - 3508 TA Utrecht, the Netherlands
	 \and
                SRON National Institute for Space Research, Sorbonnelaan 2,
                NL - 3584 CA Utrecht, the Netherlands 
         \and
                Dept. of Astronomy, University of Maryland, College Park,
                MD 20742, U.S.A.
	 \and
                NASA Goddard Space Flight Center, Code 662, Greenbelt,
                MD 20771, U.S.A.
         \and
                Istituto di Astrofisica Spaziale (CNR), Area Ricerca Roma Tor
                Vergata, Via del Fosso del Cavaliere, I - 00133 Roma, Italy
         \and
                XMM-Newton Science Operations Center,
                Vilspa Satellite Tracking Station, Apartado 50727,
                28080 Madrid, Spain
	}

\date{Received, accepted }

\abstract{ \bron\ is an X-ray transient in the Galactic bulge that was
active from January through May of 2001 and whose maximum 1-200 keV
luminosity was measured to be $5~10^{-9}$~\ecs\ which is less than
$\sim$25\% of the Eddington limit, if placed at a distance equal to
that of the galactic center. We study the X-ray data that were taken
of this moderately bright transient with instruments on BeppoSAX and
RXTE. The spectrum shows two interesting features on top of a
Comptonized continuum commonly observed in low-state X-ray binaries: a
broad emission feature peaking at 7 keV and extending from 4 to 9 keV,
and a soft excess with a color temperature below 1 keV which reveals
itself only during one week of data. High time-resolution analysis of
412 ksec worth of data fails to show bursts, coherent or
high-frequency quasi-periodic oscillations. Given the dynamic range of
the flux measurements, this would be unusual if a neutron star were
present.  \bron\ appears likely to contain a black hole. No quiescent
optical counterpart could be identified in archival data within the
5\arcsec-radius XMM error circle, but the limits are not very
constraining because of heavy extinction ($A_{\rm V}=16$).
 }

\maketitle 

\section{Introduction}
\label{intro}

Dedicated monitoring programs with the BeppoSAX Wide Field
Cameras (WFCs; e.g. In 't Zand 2001) and the RXTE Proportional Counter Array
(PCA; Swank \& Markwardt 2001) have
revealed in recent years tens of faint X-ray binary transients in a
$\sim20^{\rm o}$ field around the galactic center. 
About three quarters of these exhibit type-I X-ray bursts,
diagnosing the compact component as a neutron star. Others involve optically
confirmed black hole systems (e.g., SAX~J1819.3-2525=XTE~J1819-253
[Orosz et al. 2001]) and microquasar-type suspected black hole systems
(e.g. XTE~J1748-288 [Naik et al. 2000; Revnivtsev et al.
2000]).

A recent addition to the group of faint transients is \bron\ which was
discovered in an observation with the BeppoSAX Wide Field Cameras
during Feb. 8.8-11.5, 2001 when the average flux was 55~mCrab in the 2
to 9 keV bandpass (In~'t~Zand, Kaptein \& Heise 2001). Its galactic
coordinates are $l^{\rm II}=348\fdg4, b^{\rm II}=+0\fdg8$. This source
was followed up with sensitive X-ray devices on BeppoSAX, RXTE, and
XMM-Newton. Wijnands \& Miller (2002) reported on an analysis
carried out on part of the RXTE-PCA data and found that the soft/hard
state behavior is decoupled from the X-ray luminosity and possibly the
mass accretion rate.  They find that these data, including a detected
quasi-periodic oscillation, are inconclusive as to the nature of the
compact object.

In the present paper we discuss all the X-ray data obtained with
BeppoSAX and RXTE, and the position as determined with XMM-Newton. We
start in Sect.~\ref{obs} with an overview of all observations and the
technical details of the analysis. This is necessarily a long overview
because most observations are complicated due to source confusion.  We
continue with discussions of the 1-200 keV spectrum (Sect.~\ref{broad}),
the 3-20 keV spectral variability (Sect.~\ref{spvar}), the light curve
(Sect.~\ref{lc}), the 3-20 keV rapid variability (Sect.~\ref{var}),
and the archival search for an optical counterpart (Sect.~\ref{pos}). 
In Sect.~\ref{discussion} we discuss the implications of the measurements
for our understanding of this object. Finally, we summarize our conclusions
in Sect.~\ref{conclusions}.

\begin{table}[t]
\caption[]{Log of X-ray observations on \bron\label{tablog}}
\begin{tabular}{rll}
\hline
Day in 2001$^\dag$ (UT) & Instrument$^\ddag$ & ObsID \\
\hline
 36.557-- 39.658 & WFC           & OP10695,10696,10697 \\
 39.685-- 42.000 & WFC           & OP10703,10704 \\
 45.634-- 47.518 & WFC           & OP10733,10734 \\
 45.704-- 45.713 & PCA (3)       & 50138-01-01 \\
 47.68 -- 48.57  & NFI           & 21286001 \\
 47.729-- 47.769 & PCA (3)       & 50138-01-02 \\
 50.845-- 50.884 & PCA (2)       & 50138-01-03 \\
 56.836-- 56.856 & PCA (4)       & 50138-01-04 \\
 61.93 -- 62.09  & XMM$^\P$      & 0135520401 \\
 62.460-- 62.497 & PCA (3)       & 50138-01-05 \\
 68.836-- 68.862 & PCA (3)       & 50138-01-06 \\
 75.759-- 76.637 & WFC           & OP10898 \\
 77.247-- 77.264 & PCA (4)       & 60407-01-01 \\
 80.196-- 80.876 & WFC           & OP10927 \\
 86.007-- 86.023 & PCA (2)       & 60407-01-02 \\
 92.770-- 92.798 & PCA (2)       & 60407-01-03 \\
 93.482-- 94.438 & WFC           & OP11033 \\
 99.651-- 99.719 & PCA (3 and 2) & 60407-01-04 \\
107.673--107.712 & PCA (3)       & 60407-01-05 \\
112.797--113.586 & WFC$^\ast$    & OP11170 \\
114.049--114.079 & PCA (3)       & 60407-01-06 \\
121.745--121.770 & PCA (2)       & 60407-01-07 \\
126.974--127.002 & PCA (2)       & 60407-01-08 \\
133.923--133.950 & PCA (2)       & 60407-01-09 \\
149.635--149.665 & PCA (3)       & 60407-01-10 \\
168.441--168.482 & PCA (2)$^\ast$& 60407-01-11 \\
184.759--184.773 & PCA (3)$^\ast$& 60407-01-12 \\
313.573--313.592 & PCA (2)$^\ast$& 60407-01-13 \\
\hline
\end{tabular}

\noindent
$^\dag$ Jan 1, 2001, is MJD~51910.\\
$^\ddag$ For the PCA, the number of active PCUs outside PCU0 is given between
parentheses (for day 99 this number changed within the observation).

\noindent
$^\P$ The spectral XMM-Newton data will be dealt with in a separate paper,
the XMM-determined position is discussed in Sect.~\ref{pos}.

\noindent
$^\ast$ \bron\ was not detected in this observation.
\end{table}

\section{Observations and data analysis issues}
\label{obs}

Table~\ref{tablog} presents a log of all X-ray observations carried
out on \bron\ between 2001 Feb 5 and 2001 Nov 9.  Each of the four
data sets has a unique diagnostic value: the BeppoSAX-WFC data provide
coverage of the early parts of the outburst, the BeppoSAX-NFI data
provide simultaneous wide spectral coverage, and the RXTE-PCA data
provide coverage of the decay of the outburst combined with high time
resolution analysis capabilities. We note that in addition to the data
specified in Table~\ref{tablog}, there are daily data after Jan 22 from
the All-Sky Monitor (ASM) on RXTE.

\subsection{BeppoSAX-WFC}

\begin{figure}[t]
\psfig{figure=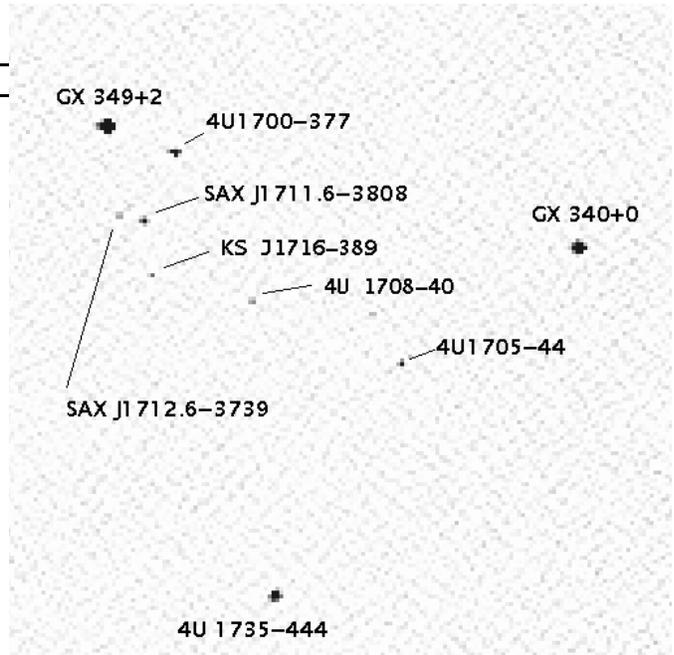,width=\columnwidth,clip=t}
\caption{Image of 13\fdg6$\times$13\fdg6 region around \bron\
taken with the WFCs on day 75 in the 2-28 keV bandpass. Pixel values
are in units of significance of detection.
\label{figwfcim}}
\end{figure}

WFC unit 1 (Jager et al. 1997) on BeppoSAX (Boella et al. 1997a)
observed the position of \bron\ seven times, at various off-axis
angles and, therefore, sensitivities in the 2-28 keV bandpass.  The
first observation started on Feb. 5 (day 36 of 2001) and the last
ended on April 23 (day 113). All observations except the last yielded
detections. The discovery was made from the second observation
(In~'t~Zand et al. 2001). The total exposure time was 320~ksec. The
statistical quality of the WFC data is limited; the highest
signal-to-noise ratio is 54 and applies to the third observation. In
all other cases this ratio is at least a factor of 2 lower. This kind
of quality only allows rough spectral
constraints. Figure~\ref{figwfcim} shows the wide-field WFC image
taken on day 75 of 2001. It shows another active source at only
33\arcmin\ from \bron.  This is \1712, a bursting low-mass X-ray
binary transient that was discovered with the WFC in 1999 (In~'t~Zand
et al. 1999, Cocchi et al. 1999). On day 75 it had an average flux of
33$\pm2$~mCrab (2-28 keV), which is similar to the peak flux in
1999. During the other WFC observations of \bron\ in 2001, \1712\ was
not detected with upper limits between 6 and 30~mCrab.  However,
archival WFC data show \1712\ sometimes slightly popping up above the
WFC detection threshold (Cocchi et al, in prep.).

\begin{figure}[t]
\psfig{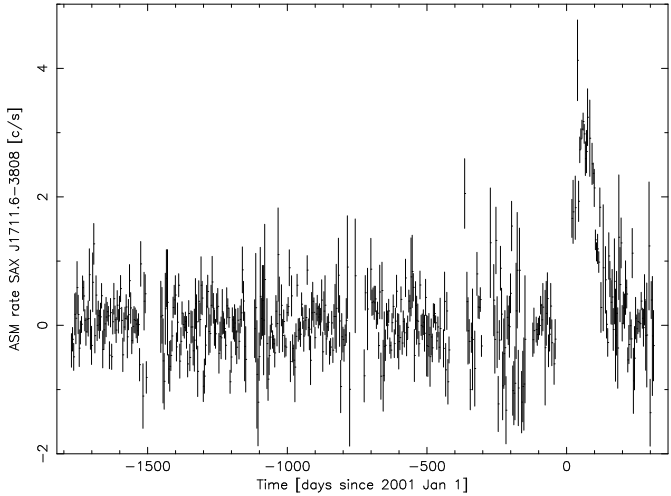}
\caption{RXTE/ASM light curve of \bron\, from 1996 Jan on,
with a time resolution of 4~d. Standard criteria were employed to exclude
bad data; most importantly data points with background levels in excess
of 10~c~s$^{-1}$. A bias level of 0.33~c~s$^{-1}$ was subtracted which
is the average level before -400 ~d.
\label{figlc1711}}
\end{figure}

\begin{figure}[t]
\psfig{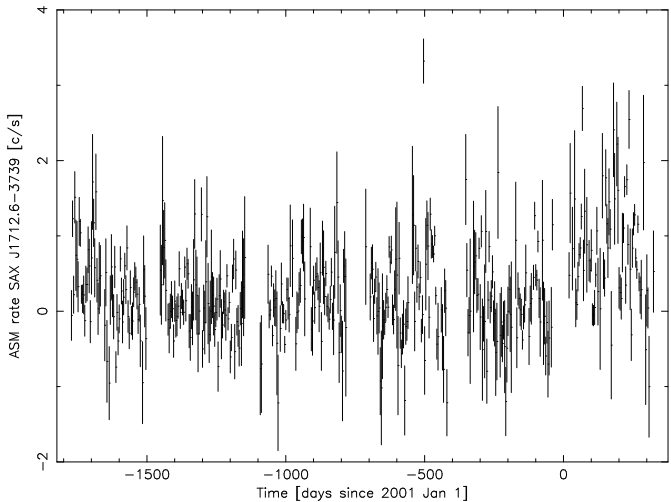}
\caption{RXTE/ASM light curve of \1712. A bias level of 0.6~c~s$^{-1}$
was subtracted.
\label{figlc1712}}
\end{figure}

\subsection{RXTE-ASM}
\label{obsasm}

ASM archival data was processed to generate a light curve of \bron\
since the start of mission operations in January 1996, see
Fig.~\ref{figlc1711}.  Only one clear outburst was detected from
\bron, in 2001. During the outburst, up to 33 ASM dwells of 90~sec
each were carried out per day, with a total number of dwells of 1194
and a total exposure time of 107 ksec. The typical sensitivity in one
dwell on a non source-confused region is 0.03 Crab units. The number 
of sources solved for per dwell is
between 7 and 37. Fig.~\ref{figlc1712} shows the ASM light curve of
\1712. The highest point here corresponds to the August 1999 outburst
when the source was discovered with WFC (In~'t~Zand et
al. 1999). Further activity is visible in 1996 and, particularly, in
2001.

\subsection{BeppoSAX-NFI}

\bron\ was observed with the Narrow-Field Instruments (NFI) on
BeppoSAX between 2001~Feb~16.68 and 17.57 UT (day 47.68-48.57 of
2001). The NFI consist of two imaging devices that operate at photon
energies below 10 keV: the Low-Energy (LECS; 0.1-10 keV bandpass;
Parmar et al. 1997) and Medium-Energy Concentrator Spectrometer (MECS;
1.6-10 keV; Boella et al. 1997b). Two other -- non-imaging -- devices
operate roughly above 10 keV: the High-Pressure Gas Scintillation
Proportional Counter (HP-GSPC; 7-34 keV; Manzo et al. 1997) and the
Phoswich Detector System (PDS; 15-220 keV; Frontera et al. 1997). The
circular fields of view of the LECS, MECS and HP-GSPC are 40\arcmin,
30\arcmin\ and 1\fdg 0 in diameter, respectively. The PDS has a
hexagonal field of view with a diameter ranging between 1\fdg 3 and
1\fdg 5. The energy resolutions at 6~keV are 8\% for the MECS (full
width at half maximum FWHM) and 11\% for the HP-GSPC. The net exposure
times are 9.5 (LECS), 36 (MECS), 36 (HP-GSPC) and 18 ksec (PDS).  The
analysis of the data sets was carried out in a standard manner:
accumulation radii of 8\arcmin\ and 4\arcmin\ were applied for the LECS
and MECS images, the normalizations of the LECS and PDS data sets
with respect to the MECS data were left as free parameters (while it
was checked that these numbers are within acceptable limits), and
spectral bins were combined up to the point where each bin contains at
least 20 photons so that the $\chi^2$ statistic applies.

The PDS consists of two pairs of detectors that each have an
independent collimator which rocks on and off the source at 192~s
intervals. The off source positions are 3\fdg5 on opposite sides from
the source. For the observation on \bron, these two positions are not
on blank sky: 2S~1711-339 is 53\arcmin\ off-axis in one off-source
pointing, while 4U~1708-40 is 50\arcmin\ in the other. On top of that,
\1712\ contaminates the signal in the on-source pointing. All three
contaminating sources have been observed on-axis with the PDS before
and the relevant data are publicly available from the BeppoSAX
archive\footnote{http://www.asdc.asi.it/bepposax/archive\_browser.html}.
2S~1711-339 is a transient X-ray source which has a fairly high X-ray
flux during 'quiescence'. It was observed during quiescence on
2000~Feb~29 (e.g., Cornelisse et al. 2002), with a 2-6 keV flux of
2.4~10$^{-11}$~\ecs, but the source was not detected above a 3$\sigma$
upper limit of 0.12 c~s$^{-1}$ in the 15-100 keV band. At other times
during quiescence, the source was seen at 2-6 keV flux levels between
0.3 and 5~10$^{-11}$~\ecs\ (Cornelisse et al. 2002), so one would
expect a maximum PDS count rate of approximately
0.24~c~s$^{-1}$. 4U~1708-40 is a persistent X-ray source with a mean
2-12 keV flux of about 10$^{-9}$~\ecs, according to publicly available
measurements with the ASM on RXTE.  A PDS observation on August 15,
1999, did not reveal a detection above a 3$\sigma$ upper limit of
0.12~c~s$^{-1}$. Judging from the ASM data, the flux is not expected
to vary by more than a factor of about 2, so also for this source we
set the upper limit to 0.24~c~s$^{-1}$.  Taking into account these
upper limits and the collimator responses, we estimate that the two
sources in the off-source pointings introduce systematic flux
uncertainties of less than one percent in the flux of \bron\ (whose
intensity is 21~c~s$^{-1}$). We verified this by comparing the PDS
spectra of both off-source pointings, for both pairs of
detectors. They fall on top of each other, and the flux does not
differ by more than 1\%.

\1712\ was observed with the PDS on August 27, 1999 (Cocchi et al., in
prep.)  when it had a 2-10 keV flux of 1.1~10$^{-10}$~\ecs\ and a
photon count rate in the PDS of 0.76~c~s$^{-1}$ (15-100 keV). During
the NFI observation on Feb 16-17, the ASM dwelled on \1712\ 18
times. The combined ASM data do not show a detection above a $3\sigma$
upper limit of approximately 3.3~10$^{-10}$~\ecs\ (2-10 keV).  During
the 2 days prior to the NFI observation of \bron, WFC observations
show that the 2-10 keV flux of \1712\ was at most equal to that during
the 1999 NFI observation.  Scaling the 1999 PDS flux to this upper
limit, we conclude that the contribution of \1712\ to the PDS flux of
\bron\ amounts to 11\% at maximum (2.3 c~s$^{-1}$), but more
likely less than 3.5\% (as inferred from the WFC observations 2 days
earlier).  In the HP-GSPC data, the contribution is expected to be
even less. We note that it is not possible to employ the PDS data in
the same way as the PCA slew data (see Sect.~\ref{obspca}) because the
on/off slews are too fast, lasting less than 3 sec.

\subsection{RXTE-PCA}
\label{obspca}

\begin{figure}[t]
\psfig{figure=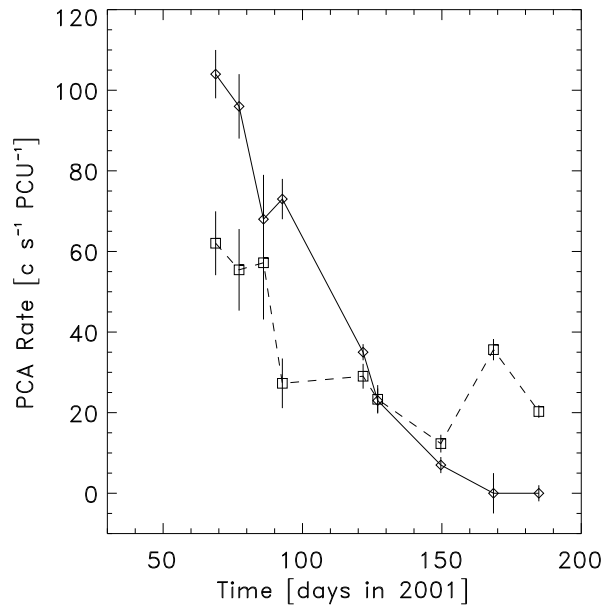,width=\columnwidth,clip=t}
\caption{2-10 keV photon rates in the top layer of all active PCA PCUs,
for \bron\ (diamonds) and \1712\ (squares)
as determined from slew data. These data represent true counting rates and
have not been corrected for collimator vignetting for either source. 
\label{figpcaslew}}
\end{figure}

RXTE observed \bron\ 19 times. The first 6 observations, amounting to
14.8 ksec exposure time, were carried out under an accepted AO5 TOO
program (PI J.  Swank), and the remainder (30.9~ksec exposure) as
public TOO observations.  Wijnands \& Miller (2002) already reported
about the first 8 public TOO observations.  We here report on all PCA
measurements. The PCA instrument (for a detailed description, see
Jahoda et al. 1996) consists of an array of 5 co-aligned Proportional
Counter Units (PCUs) that are sensitive to photons of energy 2 to 60
keV with a total collecting area of 6500~cm$^2$. The spectral
resolution is 18\% FWHM at 6 keV and the field of view is nearly
hexagonal with a size of 1$^{\rm o}$ FWHM. During the observations the
number of active PCUs varied between 1 and 5. While the first 17
observations have the same pointing (equal to the WFC-determined
position of \bron), the last two observations consisted of additional
pointings that are more than 1\degr\ from \1712, to determine the flux
of \bron\ without ambiguity at the cost of some sensitivity. During
analyses, we refrain from using PCU0, because this unit is missing its
top propane layer and hence has higher background and a different
spectral response. Furthermore, we only include PCU2 and 3, because of
calibration differences with PCU1 and 4.

\1712\ is potentially an important source of contamination in the PCA
data.  At an angle of 33\arcmin\ to the pointing axis, the collimator
response to \1712\ is 44\% with the exception of the last two observations.
To determine the contamination by \1712\ with
far better accuracy than provided by the ASM data, we investigated
data taken at times when the PCA is slewing to or from \bron,
immediately before and after each observation. The profile of the
photon count rate during these slews can be decomposed into the
contributions from both sources if the slew direction has a
sufficiently large component along the line connecting both sources,
and if both sources are steady enough during the slews to not disturb
the slew profiles in the rates. It turns out that half of all PCA
observations have useful slew data. In Fig.~\ref{figpcaslew} we
present the count rates for both sources as determined from these
slews. It is clear that \1712\ has a significant contribution which
can never be neglected: in no case is the photon rate smaller than 0.4
times that of \bron. In the final three observations it is even the sole
source of photons.

We note that the second WFC observation overlaps the first PCA
observation.  \1712\ was not detected with the WFCs and the upper
limit is relatively sensitive at 6~mCrab which translates to roughly
13 c~s$^{-1}$PCU$^{-1}$ on axis or 6~c~s$^{-1}$PCU$^{-1}$ at
33\arcmin\ off axis.

Additionally, there is contamination from the Galactic ridge. We
determined the local ridge emission from a 1~ksec exposure in the one but
last observation which was offset from \1712\ so that no emission from
\1712\ nor \bron\ was detected. The pointing is at $l^{\rm
II}=$347\fdg 69, $b^{\rm II}=$+0\fdg 84. This position is offset by
0\fdg 73 in Galactic longitude from the other pointings on \bron, but
only 0\fdg 05 in latitude (the spatial profile of the ridge is
particularly dependent on latitude).  The photon count rate derived
for the local ridge emission is 2.8 c~s$^{-1}$PCU$^{-1}$ (top layer,
3-20 keV; compare with the source photon rates in
Fig.~\ref{figpcaslew}).  The ridge emission also includes a narrow
Fe-K emission line with flux 2.1~$10^{-4}$~phot~s$^{-1}$cm$^{-2}$.

\subsection{XMM-Newton}

A public TOO was carried out with XMM-Newton from March 2.93 to 3.09
U.T. On this platform, measurements are done through 3 identical
telescopes (e.g., Jansen et al. 2001). For 2 of these, 50\% of the
X-radiation is diverted to reflection gratings. The other 50\% of the
two telescopes are collected by the EPIC-MOS1 and MOS2 CCD arrays
(Turner et al. 2001) while 100\% of the third telescope is collected
by the EPIC-pn CCD array (Str\"{u}der et al. 2001). The 3 EPIC cameras
enable imaging measurements at $\sim$6\arcsec\ resolution in a
30\arcmin-diameter field of view. The energy resolution intrinsic to
all CCDs is $E/\delta E$=20 to 50 in a 0.1 to 15 keV bandpass. The
exposure times per device are 12 ksec for EPIC-MOS1 and MOS2, and 10
ksec for EPIC-pn CCD. All cameras were operated in 'full window' data
acquisition mode. A treatment of the spectral XMM-Newton CCD data,
which is heavily piled up, and the (heavily absorbed) RGS data is deferred
to a later paper. We will here only address the source position
resulting from these data.

\section{Broad band spectrum}
\label{broad}

\begin{table}
\caption[]{Goodness of fits for 1-200 BeppoSAX spectrum.}
\begin{tabular}{lll}
Model$^\dag$ & $\chi^2_\nu$ & $\nu$ \\
\hline
wa$\times$(diskbb+pexrav)                                  & 3.703 & 116 \\
wa$\times$(diskbb+comptt)                                  & 3.609 & 116 \\
wa$\times$(diskbb+po$\times$highecut)                      & 3.453 & 116 \\
wa$\times$(pexriv)                                         & 2.538 & 117 \\
wa$\times$(diskbb+po$\times$highecut+bbody)                & 2.012 & 114 \\
wa$\times$(diskbb+pexrav+gauss)                            & 1.968 & 113 \\
wa$\times$(diskbb+pexrav+laor)                             & 1.788 & 112 \\
wa$\times$(diskbb+pexriv+gauss)                            & 1.737 & 112 \\
wa$\times$(diskbb+pexriv+laor)                             & 1.664 & 111 \\
wa$\times$(bbody+comptt+laor)                              & 1.473 & 113 \\
wa$\times$(bbody+comptt+gauss)                             & 1.458 & 113 \\
wa$\times$(diskbb+comptt+gauss)                            & 1.432 & 113 \\
wa$\times$(diskbb+comptt+laor)                             & 1.424 & 113 \\
wa$\times$(diskbb+po$\times$highecut+laor)                 & 1.386 & 112 \\
wa$\times$(diskbb+po$\times$highecut+gauss)                & 1.348 & 113 \\
wa$\times$(diskbb+po$\times$highecut+gauss)$\times$edge    & 1.308 & 111 \\
wa$\times$(diskbb+comptt+gauss)$\times$edge                & 1.308 & 111 \\
\hline
\end{tabular}
\label{tabchi}

\noindent
$^\dag$wa = interstellar absorption following Morrison
\& McCammon (1983), comptt = Comptonized spectrum,
gauss = Gauss function, edge = 1 for $E<E_{\rm e}$ and 1-exp($E/\tau$)
for $E>E_{\rm e}$, po = power law, highecut = 1 for $E<E_{\rm b}$ and
exp($-E/E_{\rm f}$) for $E>E_{\rm b}$, bbody = single-temperature black
body radiation, diskbb = disk
black body (Mitsuda et al. 1984); pexriv = Compton reflection against
an ionized medium (Magdziarz \& Zdziarski 1995); pexrav = Compton
reflection off a neutral medium (Magdziarz \& Zdziarski 1995).
\end{table}

\begin{figure}[t]
\psfig{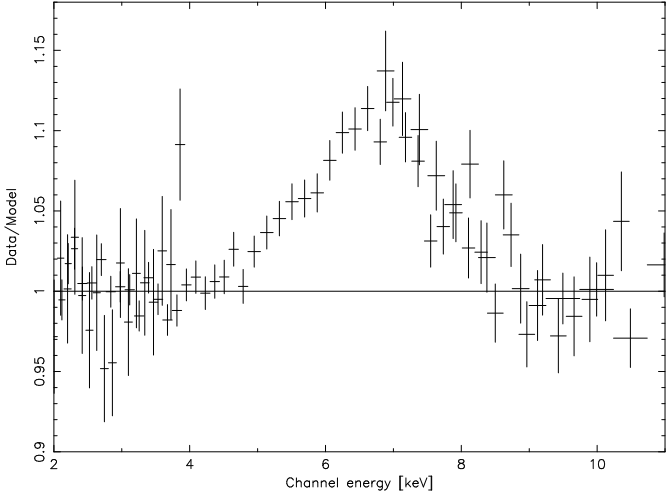}
\caption{2-11 keV NFI data of \bron divided by the best-fit Comptonized
model, excluding 4-9 keV in the fit.
\label{figratio}}
\end{figure}

The flux was fairly stable during the NFI observation; on a time scale
of an hour the variability of the source did not exceed the 5\%
level. The shape of the 1 to 200 keV spectrum can be characterized by
an absorbed power law, with a photon index of 1.7, which is moderately
cut off above 60 keV by an exponential with an e-folding energy of
130~keV.  There is no strong soft component which is typical of many
bright X-ray transients. We tested the spectrum against various
continuum models that are applicable to X-ray binaries, for instance a
(cut-off) power law, a Comptonized spectrum, bremsstrahlung radiation,
and disk black-body radiation. Data were restricted to 1-4 keV for
LECS, 1.6-10.5 keV for MECS, 7-34 keV for HP-GSPC, and 15-200 keV for
PDS. None of the continuum models adequately describe the data (see
cases in Table~\ref{tabchi} for which $\chi^2_\nu$ exceeds 2).  The
primary reason for this is a narrow-band spectral component between 4
and 9 keV which is not represented in these models. In
Fig.~\ref{figratio}, we present the residuals with respect to the
best-fit model when excluding the 4-9 keV range from the fitting. A
narrow-band emission feature is apparent, with a peak that is near the
energy expected for the Fe-K line complex.  The feature is seen in its
entirety with the MECS, and is partly confirmed by HP-GSPC data. To
accommodate the emission feature, we tested two models in combination
with a variety of continuum models. The emission-feature models
include a broad Gaussian line, optionally supplemented by an
absorption edge, and an emission line that is relativistically
broadened by Doppler shifts due to orbital motion of the emitting
material around a compact object and to gravitational redshift.  For
the latter, the model as formulated by Laor (1991) for a rotating
compact object was employed. In this model, we fixed the outer radius
at 400$R_g$ and the power-law index of the emissivity-radius
dependence at --3. None of the fixed parameters are in fact
constrained by the data. The continuum models consist of a disk black
body (according to Mitsuda et al. 1984) plus either a (high-energy cut
off) power law plus, a Comptonized thermal spectrum (Titarchuk 1994,
Titarchuk \& Lyubarskij 1995, Hua \& Titarchuk 1995) or a power-law
model reflected off cold or ionized material (following Magdziarz \&
Zdziarski 1995). The results for the goodness of fit are presented in
Table~\ref{tabchi}.  Formally, the values for $\chi^2_\nu$ are not
acceptable. For the best fit, the chance probability is as small as
about 1\%. We attribute this to calibration uncertainties. With the
introduction of a reasonable systematic flux uncertainty of 1\% per
channel, $\chi^2_\nu$ decreases to 1.0. In Table~\ref{tabfits}, we
present the parameter values for four of the best-fit models. We chose
to not consider absorption edges because the evidence for their
existence is not convincing.  Also we chose the disk black body model
for the soft excess because single-temperature black body radiation
does not fit PCA data (see Sect.~\ref{spvar}). Although one of the
four is the best fit, the other 3 are consistent at the 1 sigma level.
We conclude that 1) the fitted line width is very broad, much larger
than the spectral resolution of the MECS (33\% versus 8\% FWHM); 2)
the emission line is consistent with being symmetric with a shape that
can either be modeled with a broadened Gauss function or a
relativistically broadened emission line with a fairly high
inclination angle; 3) the absorption edge is not mandated by the data;
4) reflection models are inconsistent with the data (they particularly
have a problem with explaining the high-energy cutoff); 5) both
non-reflected continuum models give nearly equally as good a
description of the data and 6) the data does not discriminate between
single-temperature or multi-temperature black body models.

\begin{table}[tb]
\caption[]{Spectral fit parameters of the four most acceptable models to
the BeppoSAX spectrum. A systematic flux uncertainty of 1\% per channel
was added in quadrature to the statistical error.
Errors are 90\% confidence per parameter of interest (i.e.,
from range in each parameter for which $\chi^2$ is smaller than minimum
$\chi^2$ plus 2.71).}
\begin{tabular}{ll}
\hline
Model                &  wa$\times$(po$\times$highecut+diskbb+gauss) \\
$N_{\rm H}$          & $2.83^{+0.08}_{-0.09}~10^{22}$~cm$^{-2}$ \\
disk bb $kT_{\rm in}$& $0.94^{+0.27}_{-0.14}$ keV \\
photon index $\Gamma$& $1.65\pm0.01$ \\
$E_{\rm b}$          & $54.0^{+3.4}_{-3.3}$ keV\\
$E_{\rm f}$          & $127^{+10}_{-11}$ keV\\
Gauss $E_{\rm line}$ & 6.33$^{+0.24}_{-0.22}$ keV\\
Gauss line width     & 2.79$^{+0.48}_{-0.60}$ keV (FWHM) \\
Gauss line flux      & $4.8^{+1.3}_{-1.5}~10^{-3}$~phot~s$^{-1}$cm$^{-2}$ \\
$\chi^2_\nu$         & 0.999 (113 dof) \\
\hline
Model                &  wa$\times$(po$\times$highecut+diskbb+laor) \\
$N_{\rm H}$          & $2.86^{+0.04}_{-0.04}~10^{22}$~cm$^{-2}$ \\
disk bb $kT_{\rm in}$& $0.74^{+0.58}_{-0.14}$ keV \\
photon index $\Gamma$& $1.64^{+0.01}_{-0.01}$ \\
$E_{\rm b}$          & $54.5^{+3.6}_{-2.5}$ keV\\
$E_{\rm f}$          & $128^{+8}_{-10}$ keV\\
Laor $E_{\rm line}$  & 6.34$^{+0.18}_{-0.22}$ keV\\
Laor $R_{\rm in}$    & $4.5^{+1.0}_{-2.6}R_{\rm g}^\dag$ \\
Laor line flux       & $3.8^{+0.6}_{-0.4}~10^{-3}$~phot~s$^{-1}$cm$^{-2}$ \\
Laor inclination     & 86\fdg3$^{+3.7}_{-0.3}$ \\
$\chi^2_\nu$         & 0.992 (113 dof) \\
\hline
Model                &  wa$\times$(diskbb+comptt+gauss) \\
$N_{\rm H}$          & $2.65^{+0.14}_{-0.11}~10^{22}$~cm$^{-2}$ \\
disk bb $kT_{\rm in}$& $0.83^{+0.13}_{-0.23}$ keV \\
$kT$ seed photons    & $1.00^{+0.20}_{-0.38}$~keV \\
$kT$ plasma          & $25.5^{+1.5}_{-0.5}$ keV\\
optical depth        & $2.19^{+0.04}_{-0.11}$ (disk) \\
                     & $5.04^{+0.08}_{-0.23}$ (sphere  geometry)\\
Gauss $E_{\rm line}$ & $6.37^{+0.20}_{-0.19}$ keV\\
Gauss line width     & $2.58^{+0.43}_{-0.51}$ keV (FWHM) \\
Gauss line flux      & $(3.9\pm1.0)~10^{-3}$~phot~s$^{-1}$cm$^{-2}$ \\
$\chi^2_\nu$         & 1.102 (111 dof) \\
\hline
Model                &  wa$\times$(diskbb+comptt+laor) \\
$N_{\rm H}$          & $2.85^{+0.04}_{-0.06}~10^{22}$~cm$^{-2}$ \\
disk bb $kT_{\rm in}$& $0.86^{+0.09}_{-0.13}$ keV \\
$kT$ seed photons    & $<0.2$~keV \\
$kT$ plasma          & $25.7^{+1.5}_{-0.6}$ keV\\
optical depth        & $2.17^{+0.05}_{-0.10}$ (disk) \\
                     & $5.01^{+0.09}_{-0.19}$ (sphere geometry) \\
Laor $E_{\rm line}$  & $6.26^{+0.18}_{-0.10}$ keV\\
Laor $R_{\rm in}$    & $3.9^{+1.1}_{-1.9}R_{\rm g}^\dag$ \\
Laor line flux       & $4.2^{+0.5}_{-0.6}~10^{-3}$~phot~s$^{-1}$cm$^{-2}$ \\
Laor inclination     & 86\fdg5$^{+3.5}_{-0.2}$ \\
$\chi^2_\nu$         & 1.101 (113 dof) \\
\hline
\end{tabular}
\label{tabfits}

\noindent
$^\dag$$R_g=GM/c^2$
\end{table}

The disk black body accounts for 9\% of the 1-10 keV unabsorbed flux
or 3\% in the 1-200 keV band. The absorbed flux in 3 to 20 keV is
1.53~$10^{-9}$~\ecs; the unabsorbed flux in the 1 to 200 keV band is
5.0~10$^{-9}$~\ecs. The equivalent width of the broad emission feature
is between 0.35 and 0.50 keV, depending on the model.

\begin{figure}[t]
\psfig{figure=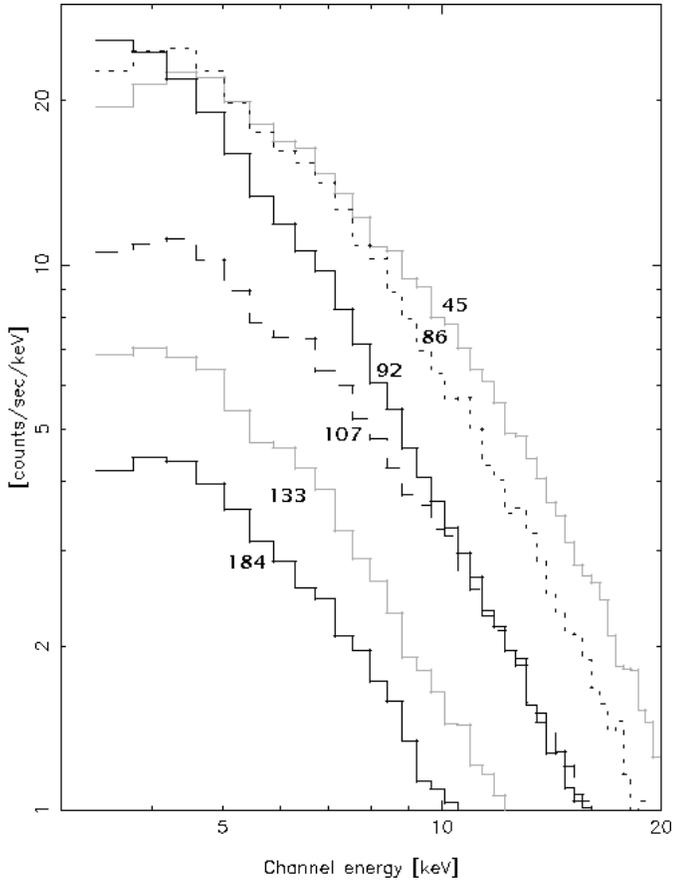,width=\columnwidth,clip=t}
\caption{Illustrative selection of 6 PCA-PCU2 spectra of \bron in 3 to 20 keV.
The number labels to each curve represent the observation day number in
2001.
\label{figpcaspectra}}
\end{figure}

\begin{figure}[t]
\psfig{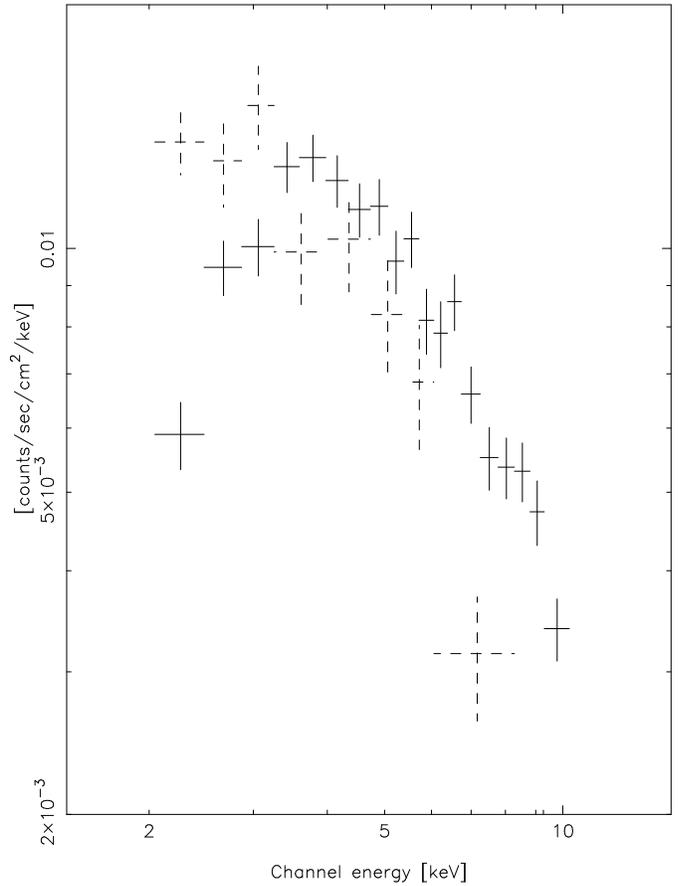}
\caption{WFC-measured spectrum of \bron between 2 and 10 keV on days 46
(solid) and 93 (dashed). 
\label{figwfcspectrum}}
\end{figure}

\begin{figure}[t]
\psfig{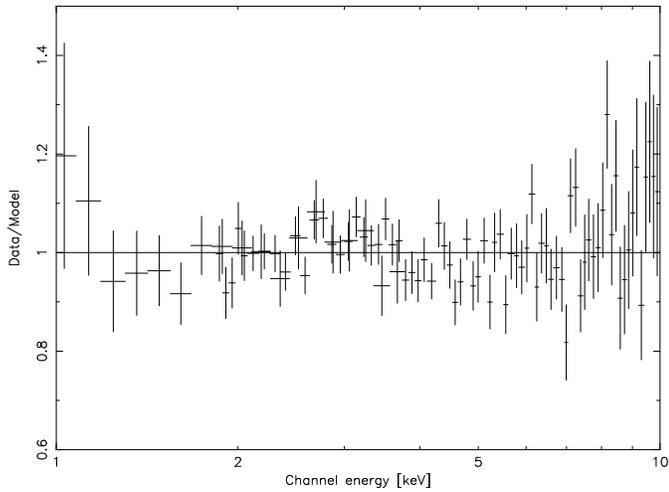}
\caption{Residuals between 1 and 10 keV of BeppoSAX NFI spectrum of
\1712\ with respect to a best-fit Comptonized spectrum with
k$T_{\rm plasma}=52$~keV, an optical depth of $0.5$ for a disk geometry,
k$T=0.04$~keV for the seed photons, $N_{\rm H}=2.08~10^{22}$~cm$^{-2}$, and
$\chi^2_\nu=1.01$ for $\nu=110$ (the data are described in detail in
Cocchi et al., in prep.). The fit was accomplished on 0.5-100 keV
data, this figure only shows part of the data. Compare with
Fig.~\ref{figratio}.
\label{fignfi1712}}
\end{figure}

\begin{figure}[t]
\psfig{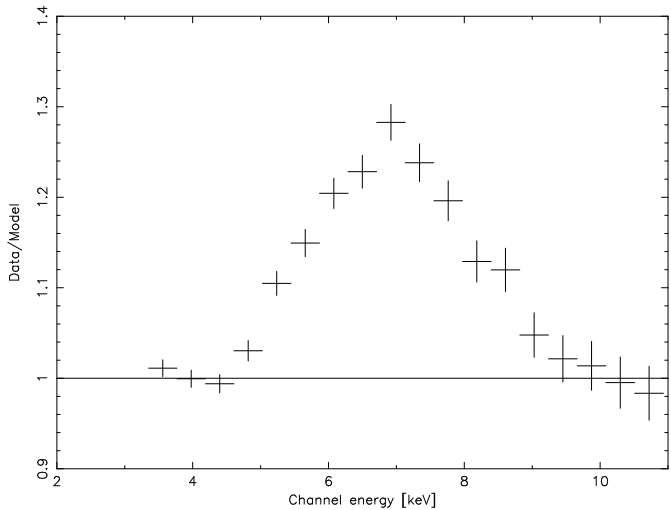}
\caption{Residual of PCA-PCU2 spectrum of \bron on day 92 with respect to an
absorbed power law. The shape of the emission feature at 6.5 keV is consistent 
with a Gaussian with a width of 2.6~keV (FWHM).
\label{figpcarat}}
\end{figure}

\begin{figure}[t]
\psfig{figure=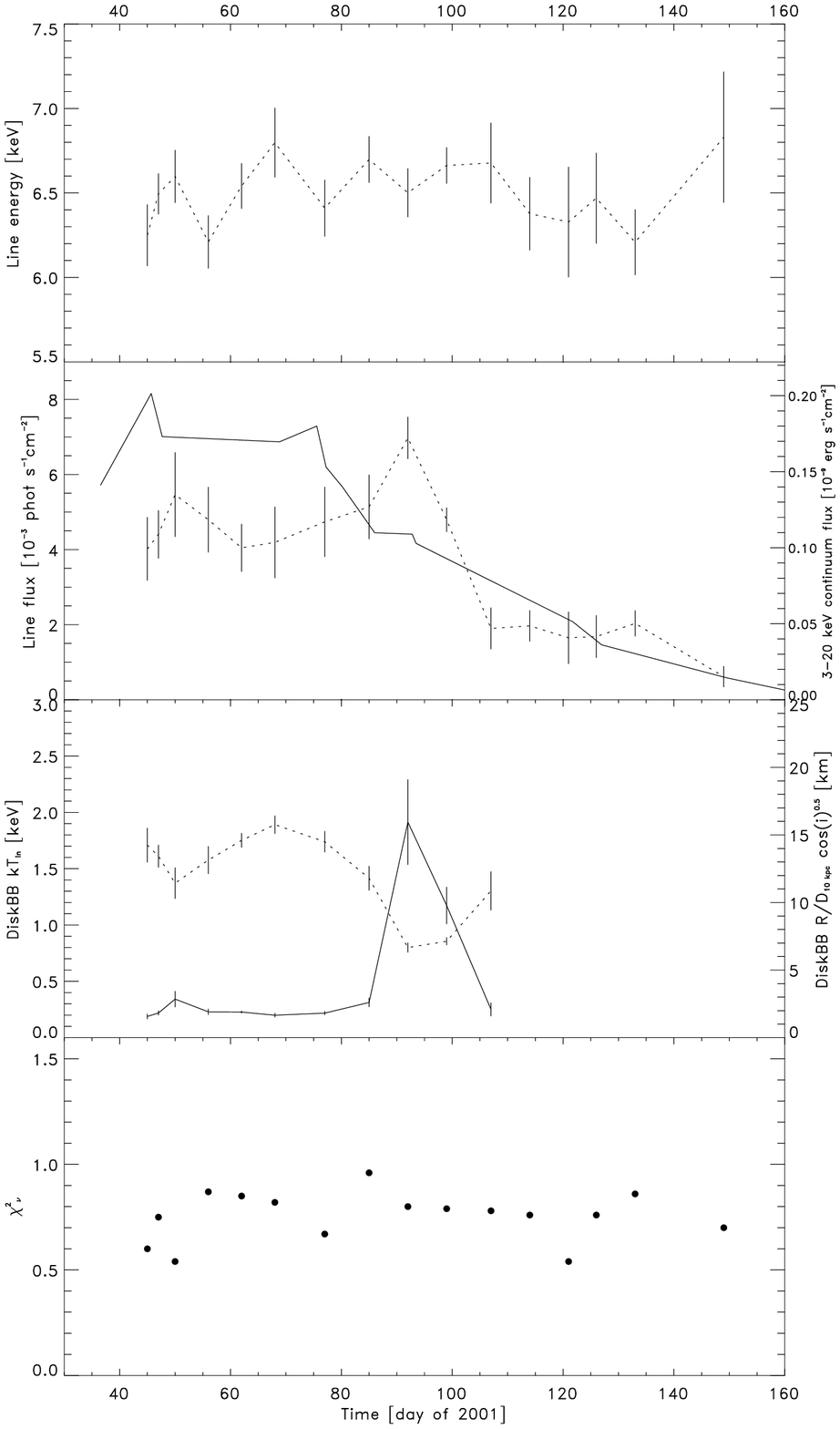,width=\columnwidth,clip=t}
\caption{Results of spectral fitting to the 16 PCA spectra. The upper
panel shows the Gaussian centroid energy, the second panel the line
flux (dashed curve) and the continuum 3-20 keV flux (solid curve). The
third panel shows the temperature at the inner edge of the accretion
disk (dashed curve) and and the radius of that edge (solid curve). The
disk black body data do not continue till the end of the outburst
because they are not constrained till then. Also, this component is
likely to be contaminated outside days 92 and 99. The bottom panel
shows the resulting reduced $\chi^2$ value per spectrum (lowest
panel).  The width of the Gaussian line was fixed at 2.6~keV (FWHM),
in accordance with the NFI spectrum and the first few PCA spectra, as
well as $N_{\rm H}$ at $2.8~10^{22}$~cm$^{-2}$.  All errors are
$\chi^2_{\rm min}+1.0$ ranges.
\label{figpcaline}}
\end{figure}

\section{Spectral variability}
\label{spvar}

Fig.~\ref{figpcaspectra} shows 6 RXTE-PCA spectra between 3 and 20 keV
for only PCU2 that illustrate the spectral changes over the course of
the outburst.  As already pointed out, there is considerable
contamination by \1712.  Only spectral features that can be attributed
to \bron\ with certainty are useful for the present discussion.This
implies that it is not prudent to discuss the broad-band continuum
from these data and we refrain from doing so.  There are two
attributable spectral features. The first is a temporary soft excess
that is strongest in the PCA observations of days 92 and 99. This is
about the time that the decay of the outburst sets in. The association
of this feature with \bron\ could be confirmed from an independent
imaging measurement with the WFC on days 93-94, see
Fig.~\ref{figwfcspectrum}.  This spectrum shows the same soft excess,
while the other 5 WFC spectra do not.  The PCA-measured excess may be
modeled by a disk black body (Mitsuda et al.  1984), with a
temperature at the inner edge of the accretion disk of
$kT=0.80\pm0.04$~keV and $0.79\pm0.02$ keV for days 92 and 99, and
inner radii of $\sqrt{{\rm cos}(i)}R_{\rm
in}/D_{\rm10~kpc}=16.5\pm1.2$ and $11.6\pm0.8$~km, respectively. A
single-temperature black body is ruled out: for the observation on day
99 a disk black body yields a best-fit with $\chi^2_\nu=1.03$ while a
single-temperature black body yields $\chi^2_\nu=3.5$ (101 dof). For all
other PCA spectra, any black body component is much fainter and of 
relatively low significance. The soft excess in the one relevant WFC spectrum
has a disk black-body temperature of $0.61\pm0.07$~keV.

The other spectral feature that can be attributed to \bron\ is the
broad emission feature at 7 keV. None of the four spectra available
for \1712\ show such a feature. These four spectra were obtained with
the PCA on August 26, 1999, with the NFI on August 27, 1999, and with
the PCA on June 17 and July 3, 2001. The last two observations were
aimed at \bron, but that source was off at that time (see
Fig.~\ref{figpcaslew}). In Fig.~\ref{fignfi1712} we show the residuals
of the NFI spectrum with respect to a Comptonized spectrum.  The
$3\sigma$ upper limit to the flux of a broad line at 6.5~keV in this
spectrum is 5.2~10$^{-4}$~phot~s$^{-1}$cm$^{-2}$.

We modeled the continuum of \bron\ with an absorbed power law, with
fixed $N_{\rm H}=2.8~10^{22}$~cm$^{-2}$ and free power law index, plus
a disk black body and included a fixed model for the Galactic
ridge. In each spectrum, the Gaussian emission line energy and flux
were left free. The width was fixed at 2.6~keV, in accordance with the
NFI spectrum. All fits are acceptable, but we note that due to the
contamination the disk black body very probably has a contribution from
\1712, as exemplified by the measurement on day 47 which is inconsistent with
that determined from the simultaneous and uncontaminated NFI measurement.
Fig.~\ref{figpcarat} shows the emission feature in the PCA data for
the observation with the highest soft excess. It is clearly present.
In Fig.~\ref{figpcaline} we plot the line energy and flux as a
function of time for all sixteen PCA observations during which \bron\
was active, as well as the disk black body parameters.  The weighted
mean of the line energy is $6.51\pm0.04$~keV. The data are consistent
with it being constant: a fit with a constant yields $\chi^2_\nu=1.11$
($\nu=15$).  The line flux decays but not exactly in tandem with the
continuum flux. The equivalent width is higher on days 85, 92 and 99,
with a maximum of 0.8~keV on day 92. This increase coincides with the
appearance of the soft excess which is modeled through the disk black
body component.

\section{Flux history}
\label{lc}

\begin{figure}[t]
\psfig{figure=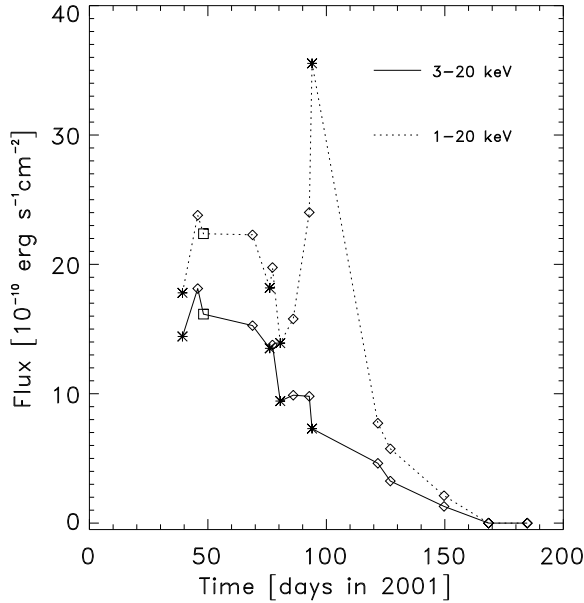,width=\columnwidth,clip=t}
\caption{The points connected through a solid line represent the history
of the unabsorbed 3-20 keV flux as measured with WFC (asterisks), NFI
(square), PCA (diamonds, only for data resolved
from \1712). The WFC data are per pointing. To calculate fluxes, spectra
were modeled with an absorbed power law plus disk black body. Subsequently,
the PCA fluxes were proportioned based on the top-layer photon rates plotted
in Fig.~\ref{figpcaslew}. Since this is not possible for the first PCA point,
we assumed a rate of  20~c~s$^{-1}$PCU$^{-1}$ for \1712\ in that case, which
equals the minimum rate over the other observations and is similar to the
upper limit in the simultaneous WFC observation. Typical flux errors are
estimated at 10\%, except for the NFI point which has an error of 2\%. 
The points connected through a dashed line represent the 1-20 keV flux, 
as estimated by a mild extrapolation of most spectra downward from 3 to 1
keV. The peak at day 90-100 is caused by the appearance of a soft excess.
%The ASM fluxes were calculated from the photon
%rates for a spectrum as measured with the NFI (absorbed power law with
%$N_{\rm H}=2.85~10^{22}$~cm$^{-2}$ and photon index 1.7).
\label{figlc}}
\end{figure}

From all X-ray observations except 8 PCA observations without useful
slew data, we generated a time profile of the unabsorbed 3-20 keV flux
as shown in Fig.~\ref{figlc}. We included the first PCA observation,
despite the absence of useful slew data, because a simultaneous WFC
observation shows \1712\ to be at least 10 times as faint as \bron\
(see Sect.~\ref{obspca}; this data point is not shown in
Fig.~\ref{figpcaslew}). At the time of discovery, the flux appears to
be on the rise to its peak of $1.8\times10^{-9}$~\ecs. However, this
is difficult to assess with certainty because the onset of the
outburst was missed. Data obtained with the ASM on RXTE are incomplete
at the time of the onset of the outburst: there is a gap in the
coverage between November 20, 2000, and January 22, 2001 (see
Fig.~\ref{figlc1711}). After January 22, the data show detections
with fluxes that are of similar magnitude as seen with the PCA, and
before November 22 there are only non-detections below a sensitivity
limit of roughly $2\times10^{-10}$~\ecs\ on a weekly time scale.
After reaching the peak flux at about the time of the NFI observation,
the source hovered at about the same flux level for one month,
Subsequently, it steadily declined by about a factor of 5 in two
months. Such a decay rate is typical for bright X-ray novae (e.g.,
Chen et al. 1997). When investigated in higher time
resolution, none of the 412~ksec worth of data show bursting activity
which would have diagnosed the compact object as a neutron star. Also,
none of the ASM data shows signs of bursting activity, but it should
be noted that the time resolution of 90~sec diminishes the sensitivity
considerably towards typical bursts with an e-folding decay time of
10~sec.

We note that the unabsorbed 3-20 keV fluxes that we find are between 25 to
48\% lower than those obtained by Wijnands \& Miller (2002), due to the
correction for the contribution by \1712.

The 3-20 keV band is not a good tracer of the bolometric flux. As spectral
analyses showed, a substantial part of the bolometric flux is
contained below 3 keV and above 20 keV. For instance, during the NFI
observation, 13\% of the unabsorbed 0.1 to 200 keV flux is contained
within 0.1-3 keV and 53\% within 20-200~keV. These percentages vary
considerably: during the PCA observation of day 92 of 2001, 74\% of the
0.1-200 keV flux is estimated to be within 0.1-3 keV. This is due to
the temporary appearance of a strong soft excess. We estimated how the
time profile of the bolometric flux might look like by extrapolating
the spectral fits that were done between 3 and 20 keV. We estimate
that the 1 to 20 keV flux is rather constant between days 45 and 70 at
a level of about $2.3~10^{-9}$~\ecs. Thereafter, it starts to decline,
but revives after 10 days to an absolute peak of 4~10$^{-9}$~\ecs\
whereafter the decline resumes. This 1-20 keV peak luminosity happens
to be close to the 1-200~keV luminosity during the NFI measurement
(5~10$^{-9}$~\ecs), when there was no strong soft excess present.

\section{Rapid variability of the continuum} 
\label{var}

\begin{center}
\begin{table*}
\caption[]{Characteristic numbers for rapid variability in RXTE data.}
\label{qpo-data}
\begin{tabular}{cccccccc}
\hline
Date    & Obs &  Total & QPO1  &  QPO1   & QPO2    & QPO2      & $f_{1/2}$ \\
(day   & Id. No. &   rms  & freq. & rms amp.& freq.   & rms amp.  &       \\
in 2001) &     &        & (Hz)  &         &  (Hz)   &           & (Hz)  \\
\hline
\hline
 45.7 & 50138-05-01 &   25\% &  ---  &   ---   &    ---  &   ---   & 0.81 \\
 47.7 & 50138-05-02 &   24\% &  ---  &   ---   &  0.81   &   5.3\% & 0.83 \\
 50.8 & 50138-05-03 &   23\% & 0.49  &  3.1\%  &   1.0   &  11.2\% & 0.41 \\
 56.8 & 50138-05-04 &   23\% &  ---  &   ---   &  0.98   &   4.8\% & 0.95 \\
 62.5 & 50138-05-05 &   22\% &  ---  &   ---   &   1.2   &   9.9\% & 0.50 \\
 68.8 & 50138-05-06 &   18\% &  1.1  &  4.1\%  &   2.0   &   6.8\% & 0.66 \\
 77.2 & 60407-01-01 &   19\% &  1.1  &  4.0\%  &   2.1   &   7.6\% & 0.72 \\
 86.0 & 60407-01-02 &   19\% &  ---  &   ---   &   2.2   &   3.1\% & 1.3 \\
 92.8 & 60407-01-03 &   13\% &  ---  &   ---   &   6.9   &   3.4\% & 2.9 \\
 99.7 & 60407-01-04 &   12\% &  ---  &   ---   &   7.3   &   1.4\% & 5.1 \\
\hline
\end{tabular}
\end{table*}
\end{center}

\def\srca{SAX J1711.6--3808}
\def\srcb{SAX J1712.6--3739}

We searched the RXTE PCA observations for both coherent pulsations and
other incoherent rapid variability such as quasiperiodic oscillations
(QPOs).  The PCA data are taken in several different modes.  For fast
timing analysis, data in the ``\verb|E_125us_64M_0_1s|'' mode were
used.  This mode provides individual X-ray tagged events with 125
$\mu$s time precision and 64 bins of energy information.

Pulsations were searched for using the Fourier Transform technique.
The PCA events were selected to be in the 2--60 keV range, and
extracted into light curve segments of 1024~s duration, sampled at a
frequency of 2048 Hz (i.e., Nyquist frequency of 1024 Hz).  Fast
Fourier transforms of each light curve were computed and the resulting
power spectra were averaged.  We searched for pulsations only from
days 45--100 of year 2001, where \srca\ was known to be relatively
active, resulting in an exposure time of 20.9 ksec.  This selection
includes XTE observation IDs 50138-05-01-00 through -06-00 and
60407-01-01-00 through -04-00.  Above a frequency of 10 Hz, the
strongest Leahy-normalized (Leahy et al. 1983) Fourier power is 5.06
at a frequency of 88.53418 Hz.  However, the 99\% detection limit for
$10^{6}$ trials is a power of 5.48, so no pulsations were detected at
99\% confidence.  The maximum power measured yields a 95\% upper limit
to pulsations of 1.5\% fractional (r.m.s.) in the 2--60 keV band.

We also searched for broader high frequency QPOs.  Oscillations in the
range 300--1300 Hz have been detected from a multitude of neutron star
low mass X-ray binaries, and can be as narrow in frequency as a few
Hertz, or as broad as ~100 Hz. Power spectra in the same energy band
were rebinned in frequency and examined for excess power.  We found no
significant oscillations.  The 95\% upper limit to the fractional
r.m.s. variations is 1.4\% for QPOs with a full width half-max
(FWHM) of 10 Hz, and 1.5\% for a FWHM of 100 Hz.

Strohmayer (2001a \& 2001b) has recently shown that in the black hole
candidates GRS 1915+105 and GRO J1655--40, significant QPOs are
present only in the X-ray band above 13 keV.  We thus also divided the
X-ray events into two bands, 2--8 keV and 8--60 keV, to search for
energy dependent phenomena.  No QPOs were detected in the soft and
hard bands, with 95\% upper limits of 2.1\% and 2.3\% fractional
r.m.s. respectively, for a hypothetical QPO with FHWM of 50~Hz.

\begin{figure}[t]
\psfig{figure=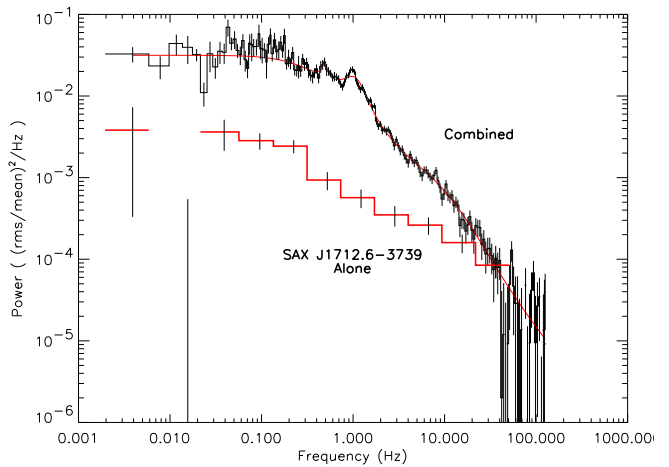,width=\columnwidth,clip=t,angle=90.}
\caption{Comparison of the XTE PCA power spectrum of MJD 51959.8 (top
curve) --- which may include flux from both \srca\ and \srcb\ --- with
the PCA power spectrum of \srcb\ alone taken in 1999 (bottom curve).
The best fitting model including continuum and QPOs is shown as a
solid smooth line. \label{pca-powspec}}
\end{figure}

Below frequencies of 10~Hz the power spectrum is dominated by
incoherent red noise (Figure~\ref{pca-powspec}; see also Wijnands \&
Miller 2002).  Generally speaking the power spectrum in the 1 mHz --
128 Hz band has a flat top with power law roll-off ($\propto f^{-1}$)
above about 1 Hz.  However, the shape of the spectrum does not
perfectly obey this prescription.  First of all there are noticeable
QPOs and harmonics present, especially in the observations that
precede day 100.  Also, the power law decline above the cut-off is not
a perfect power law but often has a smooth shelf, which is mildly
visible in Figure~\ref{pca-powspec}.

Because \srcb\ is also present in the field of view, it potentially
contaminates the power spectrum of \srca.  We have examined the power
spectrum of \srcb\, both in the last few observations of this
observing program where \srca\ was known to be quiescent, and in
observations from August 1999 (Observation IDs 40428-01-01-00 and
-01).  In those observations \srcb\ generally appeared to have weak
variability, on the order of 8\% total fractional r.m.s. fluctuations
over the entire 0.001--128 Hz band.  The average power spectrum of
\srcb\ from the 1999 PCA observations is shown in the bottom portion
of Figure~\ref{pca-powspec}.  It can be seen that the variability of
\srcb\ is weak compared to the 20\%--25\% variability of \srca\ during
the peak of its outburst, but \srcb\ may indeed contribute to the
``shelf'' seen around 10 Hz.

In principle the power spectrum should be dominated by the variability
of \srca.  Unfortunately, the flux of \srcb\ in the time range of
interest is a factor of $\sim$2 greater than has been seen before by
the PCA.  Thus, we have no direct experience regarding the variability
of \srcb\ as it existed during the peak of the outburst of \srca.  As
such we will avoid presenting detailed results on the noise continuum
and focus on the low frequency QPO features.  The continuum model
employed consisted of two additive continuum components of the form
$A/(1+(f/f_o)^2)^{\alpha/2}$ where $A$ is the normalization and
$\alpha$ is the asymptotic power law index above the break frequency
$f_o$, plus a constant to represent the Poisson noise level.  The two
additive components were used in order to capture both the main break
in the power spectrum and the ``shelf.''

The QPOs are seen between 0.5 and 2.5 Hz, near the break in the noise
continuum.  When two QPOs are formally detected, they appear to have a
1:2 harmonic relationship, dominated primarily by the 2nd harmonic.
Even when the first harmonic is not detected (owing in part to
imperfections in the continuum model) there appears subjectively to be
some excess noise at the first harmonic position.  The centroid
frequencies and amplitudes of the first and second harmonics are
presented in Table~\ref{qpo-data} as QPO1 and QPO2 respectively.  The
``Total rms'' column refers to the total fractional r.m.s. variability
in the observation in the 0.001--128 Hz frequency band.  Where a
non-detection is listed in Table~\ref{qpo-data}, we refrain from
placing upper limits given the above discussions.  The FWHM of the
lower frequency QPO ranged from 0.06 to 0.36 Hz, while that for the
upper peak ranged from 0.08 to 0.95 Hz.  The total variability
declined as the X-ray flux of \srca\ declined, and during that time
the frequencies of the QPOs were seen to increase slightly.

The ``break'' frequency of the continuum level is also indicated in
Table~\ref{qpo-data}.  The tabulated value is $f_{1/2}$, the frequency
at which the continuum reaches half its maximum value, for the larger
of the two continuum components.  This quantity is comparable to the
half-width at half-maximum value often quoted when authors fit a
zero-centered Lorentzian to the low frequency continuum component of a
power spectrum.

\begin{figure}[t]
\psfig{figure=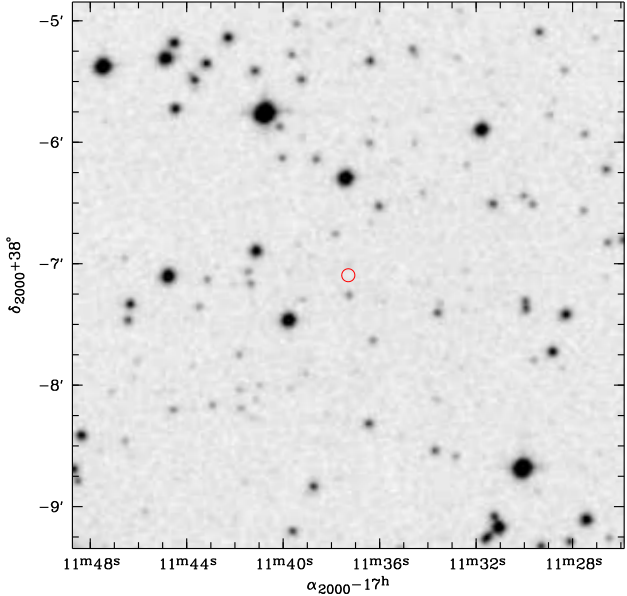,width=\columnwidth,clip=t,angle=270}
\caption{DSS sky image in $R$ around XMM-Newton position of \bron\ (circle).
The limiting magnitude is $R=22.0$.
\label{figoptical}}
\end{figure}

The nearby source \srcb\ contributes an appreciable fraction of the
total flux observed by the PCA.  For PCA observations pointed directly
at \srca\, the collimator response to \srcb\ is approximately 44\%.
For times before day 100 where it was possible to estimate the fluxes
of both sources independently using the on- or off-source PCA slews,
\srcb\ contributed between 27\% and 46\% of the total detected flux,
with an average of 37\%.  Thus, if one were to consider the QPOs to
come from \srca\ alone, the fractional r.m.s. upper limits and values
quoted above and in Table~\ref{qpo-data} are slight underestimates and 
should be revised upward by approximately 25\%.

Wijnands \& Miller (2002) also presented power spectral parameters of
\srca\, derived from PCA observations in the 60407 series.  Where QPOs
are detected both in Table~\ref{qpo-data} and in Wijnands \& Miller
(2002), there is a reasonably good agreement between in the two
values.  The values for the break in the continuum differ by a factor
of $\sim$2.2, which is not unreasonable because the continuum models
were quite different: Wijnands \& Miller (2002) use a broken power law.
However, we urge extreme caution in interpreting the later
observations from the PCA, which are the most likely to be
contaminated by \srcb.  For example, observations on days 92.8 and
99.7 have r.m.s. variabilities of 13\% and 12\% respectively.  These
are quite close to the $\sim$8\% variability seen from \srcb\ alone.
We also see a general positive trend between the QPO frequency and the
break frequency.

\section{Limits on the optical counterpart}
\label{pos}

The XMM-Newton EPIC data yield the most accurate position for \bron:
$\alpha_{2000.0}=17^{\rm h}11^{\rm m}37.1^{\rm s}$,
$\delta_{2000.0}=-38^{\rm o}07\arcmin 5\farcs7$ with an error radius
of 3\farcs2 (90\% confidence). This is 1\farcm4 from the WFC-determined
position
(In~'t~Zand et al. 2001). In Fig.~\ref{figoptical} the XMM-Newton
error circle is plotted on an image in $R$ from the Digitized Sky
Survey (DSS, 2nd generation). No candidate optical counterpart is
visible. The limiting magnitude of this image is $R=22.0$.

For $N_{\rm H}=2.85\times10^{22}$~cm$^{-2}$, Predehl \& Schmitt (1995)
predict 15.9 magnitudes of visual extinction. The extinction in the
$R$ band is expected to be 11.9 magnitudes (employing
$A_R/A_V=0.75$). This brings down the effective limiting magnitude to
a moderate $R=10.1$, or $M_R=-4.5$ for a canonical 8.5~kpc
distance. Thus, the lack of an optical counterpart in the DSS image is
not very constraining. A B0(V)-type star with $R=-4.0$ (e.g., Drilling
\& Landolt 2000) would have been detected if it were closer than
6.5~kpc. If such a star would have evolved off the main sequence into
a giant, the limiting distance would be 20 kpc. For later stellar
types the limiting distances are smaller, being roughly 1 kpc for a
B8(V) star. We conclude that it is slightly more likely that the
companion star is a late type star. To find an optical counterpart, it
is most opportune to search in the infrared, for instance in the $K$
band. The extinction in this band is only 1.8 mag (employing
$A_K/A_V=0.112$ following Rieke \& Lebofsky 1985). A K0 star, typical
for a quiescent low-mass X-ray binary (e.g., van Paradijs \&
McClintock 1995), would have $M_K=3.6$ (for $V-K=2.31$, following
Bessell \& Brett 1988). At a distance of 8.5~kpc this would imply
$K=20.1$. We note that another argument in favor of a low-mass X-ray
binary (LMXB) hypothesis is that most transients in this field are
LMXBs (e.g., In 't Zand 2001).

\section{Discussion}
\label{discussion}

The uncertain distance makes it difficult to assess the peak
luminosity accurately, but the low galactic latitude places \bron\ in
the galactic disk and diminishes the likelihood of large distances
through association with the Galactic halo. For a canonical distance
of 8.5~kpc the 1-200 keV unabsorbed luminosity would be
$(4\pm1)~10^{37}$~\lum. For masses in excess of 1.4~M$_\odot$ the
implied luminosity is below 25\% of the Eddington limit. The source
would have to be at a distance at least twice that to the Galactic
center for the peak luminosity to be near Eddington.  Despite the lack
of coverage of the onset, the available data suggest that the true
peak flux of the outburst was not much higher than observed because
the source flux remained on a plateau for a month after the discovery
before it started to decay.

Wijnands \& Miller (2002) announce a decoupling between the 3 to 20
keV luminosity and the state transitions in the sense that the soft
component appears at intermediate luminosities instead of at the peak
as would be expected. They consider it unlikely that this may be
explained by the evolution of the flux above 20 keV and conclude that
the decoupling is a significant physical effect. However, the picture
becomes different when the flux {\em below} 3 keV is also taken into
account. Our analysis shows (Fig.~\ref{figlc}) that there is an
important bolometric correction from below 3 keV. In fact, the soft
excess contains so much flux that its appearance marks the peak flux
for the whole outburst in the 1--20 keV band. While keeping in mind
that a substantial fraction of the bolometric flux may be outside the
1-20 keV band, as evidenced by the NFI data, the conclusion is
justified that there is no hard evidence for a decoupling of the
bolometric flux and state transitions.

The 6~keV emission feature in \bron\ is characterized by 1)~a centroid
energy of 6.3-6.7~keV; 2)~a FWHM of about 2.6 keV and an equivalent
width of 0.3 to 0.8~keV; 3)~little evidence for asymmetries or
structure; 4)~no
evidence for line shape variability during the outburst when the line
flux declines by one order of magnitude; and 5) temporary factor-of-2
to 3 increase in the equivalent width simultaneous with a strong
increase of a disk black body component at substantially lower
energies.

The broadness is exceptional though not unprecedented among neutron
star low-mass X-ray binaries.  Asai et al. (2000) recently analyzed
Fe-K line emission properties of 20 such LMXBs as measured with ASCA
with relatively high energy resolution (2\% at 6 keV for SIS and 8\%
for GIS), and found such lines in half of all cases.  The FWHM in
those detections ranges up to 0.7~keV. This is a confirmation of
earlier EXOSAT work compiled by Gottwald et al. (1995) who find widths
of up to $\sim1$~keV. Recently, two cases were reported of Fe-K
emission lines in neutron star systems with FWHM in excess of 1 keV:
Ser X-1 with a FWHM of $2.3\pm0.33$~keV (Oosterbroek et al. 2001), and
GX~354-0, with a width of $3.0\pm0.7$~keV (Narita et al. 2001).

While unusual for NS systems, the line width has relatively more
precedents among BH systems. There are at least 7 cases: the transient
micro-quasar XTE~J1748-288 (Miller et al. 2001), the transients
GX~339-4 (Feng et al. 2001), XTE~J1550-564 (Sobczak et al. 2000),
GRO~J1655-40 (Tomsick et al. 1999), 4U~1543-47 (van der Woerd et
al. 1989) and XTE~J2012+381 (Campana et al. 2002), and the
persistently bright Cyg X-1 (Barr et al. 1985; Fabian et al. 1989;
Miller et al.  2002).  A recent BeppoSAX observation of Cyg X-1
(Frontera et al. 2001) in its low/hard state shows a Gaussian profile
centered on 6.2~keV with a FWHM of 2.9~keV and an equivalent width of
0.35~keV. These values are similar to those for \bron. The width and
prominence of the line in Cyg X-1 are less when the source is in the
soft state.

A black hole nature of \bron\ would also be consistent with the lack
of coherent oscillations in the PCA data and the lack of type-I X-ray
bursts in all X-ray data. The total exposure time on the active \bron\
is about 412~ksec. For the exhibited range of luminosities (assuming a
canonical distance of 8.5~kpc), this should have been ample
opportunity to detect a type-I burst if the compact object would have
been a neutron star (e.g., In 't Zand 2001). Of all 24 X-ray binary
transients within 20$^{\rm o}$ from the Galactic center
that WFC detected persistent emission from,
seven failed
to exhibit type-I X-ray bursts: GRO~J1655-40, GRS~J1737-31, GRS 1739-278
XTE~J1748-288, XTE~J1755-324, SAX~J1819.3-2525 and SAX~J1711.6-3808. Apart
from the latter, these are
either dynamically confirmed black hole candidates (Bailyn et al. 1995,
Orosz et al. 2001) or suspected black holes on arguments different from
the lack of type-I bursts (see Cui et al. 1997 for GRS J1737-31, Borozdin et
al. 1998 for GRS 1739-278, e.g. Miller et al. 2001 for XTE~J1748-288, and
Goldoni et al. 1999 for XTE~J1755-324).

It is tempting to explain the line broadness as relativistic
broadening, in analogy to such features in a number of active galactic
nuclei, most notably in MCG-6-30-15 by Tanaka et al. (1995) and
NGC~3516 by Nandra et al.  (1999; for a recent review, see Fabian
2001). The line energies in those cases are about 6.5~keV, and the
equivalent widths between 0.3 and 0.6~keV.  If we do this, our data
suggests that we are viewing this system nearly edge on. If the
inclination angle is really that high, one would have expected orbital
signatures ([partial] eclipses, dips) in the light curve, but none
were detected. This may have been chance coincidence, since our data
coverage is not exhaustive (i.e., 4.7 days net exposure time which
compares to BHC orbital periods of 0.2 to 6.5~d).

In analogy to other Galactic stellar black hole X-ray transients, the
relatively low luminosity, the hard continuum spectrum in combination
with a weak black body component, and the power density spectrum show
that this source was primarily observed in the so-called low/hard
state (e.g., Tanaka \& Lewin 1995) when the inner edge of the
optically thick accretion disk is thought to be much further out than
the innermost stable circular orbit (6$R_g$ for a non-spinning compact
object). Therefore, if the broadening of the Fe-K line is relativistic
in nature, either it is unrelated to the\ accretion disk or the
low/hard state is unrelated to the radius of the inner edge of the
accretion disk. Both these possibilities seem unlikely.

An alternative to the relativistic interpretation of the line
broadening is Comptonization of the line photons by the $kT_{\rm
e}=26$~keV plasma that is measured through the continuum. The expected
width is $\sigma_E/E=E\tau^2/m_{\rm e}c^2$. The width is consistent
with the measured value if $\tau=3.8$. This value is in rough
agreement with the measurements. Broadening due to Comptonization
seems more likely than due to relativistic orbital motion because it
would naturally explain the line flux increase at the time of the soft
excess increase. When a soft excess appears, the inner edge of the
accretion disk is thought to move closer in (due to an increased mass
accretion rate). As a result, the emission from the disk becomes
harder. This may yield higher Fe-K fluxes. The lack of change in line
energy and broadness (compare Figs.~\ref{figratio} and
\ref{figpcarat}) indicates that the plasma cloud is not related to the
accretion disk but located in a spherical geometry around the compact
object.

Another alternative is Comptonization in approaching and receding
outflows. Fender (2001) shows that low/hard state black hole X-ray
binaries like 1E~1740.7-2942 and Cyg X-1 exhibit radio emission that
is thought to arise from collimated outflows which is supported in
some cases by spatially resolved observations. A correlation between
hard X-ray and radio flux suggests that these outflows do not have
large bulk Lorentz factors, unless both fluxes are beamed by the same
factor.  This is in line with the width of the Fe-K feature in \bron\
which would imply Lorentz factors up to at most 2.

\section{Summary}
\label{conclusions}

\bron\ is a transient X-ray source which was seen active for the first
time in January through May 2001. It was moderately bright, indicating
sub-Eddington accretion levels throughout the outburst. The Galactic
latitude, brightness and spectrum identify \bron\ as a Galactic X-ray
binary; optical confirmation proves difficult because of the large
extinction. The lack of type-I X-ray bursts and coherent oscillations,
and the exhibition of a broad Fe-K emission feature suggest that the
primary in \bron\ is a black hole. 

\acknowledgement
We thank Tim Oosterbroek, Takamura Tamura, Carlo Ferrigno,
and Keith Jahoda for assistance in the various aspects of the data analysis,
Jelle Kaastra for useful discussions, and Ron Remillard for his help in
obtaining the ASM light curves. JZ and EK acknowledge financial
support from the Netherlands Organization for Scientific Research (NWO).

\end{document}